\documentclass[letterpaper, 10 pt, conference]{ieeeconf}  
\IEEEoverridecommandlockouts                              
\title{\LARGE \bf
A self-paced BCI system with low latency for motor imagery onset detection based on time series prediction paradigm
}
\author{Navid Ayoobi$^{*}$ and Elnaz Banan Sadeghian$^{*}$
\thanks{$^{*}$Navid Ayoobi and Elnaz Banan Sadeghian are with the Department of Electrical and Computer Engineering,
        Stevens Institute of Technology, Hoboken, USA.
        {(E-mail: \tt\small  nayoobi@stevens.edu,}
        {\tt\small ebsadegh@stevens.edu})
        }%
}
\usepackage{cite}
\usepackage{amsmath,amssymb,amsfonts}
\usepackage{algorithmic}
\usepackage{graphicx}
\usepackage{textcomp}
\usepackage{longtable}
\usepackage{multirow}
\usepackage{xcolor}
\usepackage{color, colortbl}
\definecolor{Gray}{gray}{0.92}
\definecolor{DarkGray}{gray}{0.8}
\usepackage{caption}
\def\BibTeX{{\rm B\kern-.05em{\sc i\kern-.025em b}\kern-.08em
    T\kern-.1667em\lower.7ex\hbox{E}\kern-.125emX}}
    
\begin{document}
\maketitle
\thispagestyle{empty}
\pagestyle{empty}

\begin{abstract}

In a self-paced motor-imagery brain-computer interface (MI-BCI), the onsets of the MI commands presented in a continuous electroencephalogram (EEG) signal are unknown.
To detect these onsets, most self-paced approaches apply a window function on the continuous EEG signal and split it into long segments for further analysis.
As a result, the system has a high latency.
To reduce the system latency, we propose an algorithm based on the time series prediction concept and use the data of the previously received time samples to predict the upcoming time samples.
Our predictor is an encoder-decoder (ED) network built with long short-term memory (LSTM) units.
The onsets of the MI commands are detected shortly by comparing the incoming signal with the predicted signal.  
The proposed method is validated on dataset IVc from BCI competition III.
The simulation results show that the proposed algorithm improves the average F1-score achieved by the winner of the competition by 26.7\% for latencies shorter than one second.
\end{abstract}

\section{INTRODUCTION}

Individuals with physical disabilities can benefit from brain-computer interface (BCI) systems to move and communicate with their surroundings using only their thoughts \cite{brusini2021systematic}. 
One of the methods to acquire the electrical activities of the brain is electroencephalography.
Electroencephalogram (EEG) signals are widely used in BCI applications because of their non-invasiveness and low cost recording \cite{hernandez2021toward}.
The user's imagination of limb movement can be captured as motor imagery (MI) patterns in EEG signals.
These signals are then interpreted as commands by the BCI system. 
The MI patterns defined as the commands in a BCI system are called MI-task signals, for example, the imagination of the left-hand and right-hand movements.
Most current BCI studies focus on classifying the MI-task signals with predetermined onset and end points.
These techniques are categorized as synchronous or cue-paced BCI \cite{sadeghian2008combining}.
In real-world applications, however, the BCI system must analyze continuously incoming signals where the onsets and durations of MI-task signals and rest states are unknown.
The BCI systems analyzing these signals are categorized as asynchronous or self-paced BCI \cite{sadeghian2007continuous, sadeghian2008fractal, sadeghian2008erders}.
A self-paced BCI system must have low latency and a quick reaction to the changes in user brain patterns in order to be utilized in real-world applications.

Most current self-paced approaches apply a time window on the continuous EEG signal to split it into shorter segments for further analysis \cite{kus2012asynchronous,hsu2011continuous,an2016design}:
Kus \textit{et al.} segment the continuous EEG signal into $2$-second intervals and extract spectral power in several
frequency bands as features from spatially filtered signals \cite{kus2012asynchronous}.
The extracted features are then fed to a multinomial logistic regression classifier for detecting the MI-task signals.
Hsu \cite{hsu2011continuous} establishes a self-paced BCI system based on a two-stage recognition procedure where in the first stage, the EEG signal is segmented into $1$-second intervals, and in the second stage, multiresolution fractal features are extracted to feed an SVM classifier in order to discriminate the left and right finger lifts.
An et al. \cite{an2016design} design a self-paced BCI system for controlling a virtual Avatar using four MI commands where the continuous EEG signal is segmented into smaller intervals by applying a sliding window of duration $3$ seconds. 
The filter-bank common spatial pattern (FBCSP) algorithm is then used to extract features in each interval, and a regularized discriminant analysis classifier is adopted to distinguish different MIs.
However, all the above approaches introduce high latencies due to the wide window functions utilized to segment the continuous signal.

A secondary challenge in designing a self-paced BCI system is the detection of rest-state onsets in a continuous EEG signal due to the diverse patterns of the rest states that may be present in the signal.
The reason is that the rest states include both idling states, where the user is not imagining a limb movement, and MI signals that are not defined as a command in the BCI system \cite{zhang2007algorithm}.
Consequently, it is difficult to train a classifier based on a limited number of rest state patterns provided in the training set that can also identify previously unseen forms of rest states that may occur in the test stage.

The present study uses the time-series prediction concept to develop a self-paced BCI method that has low latency and needs no rest-state trials in the training phase.
The proposed method is a deep predicting encoder-decoder network trained exclusively on the MI-task signals.
The predicting network processes the received EEG signal and forecasts the future time samples of the upcoming signal. 
When the next time samples corresponding to the predicted ones are received, the onset of an MI-task signal can be detected by comparing these two sequences.
The length of the predicted sequence can control the system latency.
The suggested approach is tested on segments with length of $0.25s$, $0.5s$, $0.75s$ and $1s$.
In all cases, our method outperforms the algorithm proposed by the winner of BCI competition III and improves the average F1-score by $26.7\%$ compared to the score gained by the winner.

The rest of the article is organized as follows.
The dataset used and preprocessing steps are presented in section \ref{section_data}.
The proposed method is elaborated in section \ref{section_method}.
The results are presented and discussed in Section \ref{section_results}.
Finally, section \ref{conclusion} concludes this study.

\section{Dataset and preprocessing}\label{section_data}
\subsection{Dataset} 

In this work, we use the training set of dataset IVc of BCI competition III collected from one healthy individual \cite{dornhege2004boosting}. 
The subject performed left hand and right foot motor imageries.
We use $70\%$ of the training set of the dataset as our training trials and the remaining $30\%$ for the testing.
We added the rest-state trials from the testing set of the original dataset with random durations and positions to our MI-task trials to produce a continuous signal for our testing stage.

\subsection{Preprocessing}

\subsubsection{Laplacian and bandpass filtering}
Two noise sources for EEG signals are non-EEG artifacts such as eye movements and non-mu EEG components. We use small Laplacian filtering \cite{mcfarland1997spatial} to remove artifacts and improve the signal-to-noise ratio.
We also use a bandpass filter with cutoff frequencies of 6 and 13 Hz to extract the most significant event-related synchronization (ERS) and event-related desynchronization (ERD) patterns from our dataset \cite{pfurtscheller1999event}. 

\subsubsection{Dimensionality reduction} 

We apply principal component analysis (PCA) \cite{wold1987principal} to the MI-task trials and employ the first $m^{\prime}$ principal components to project the data. Hence, the number of channels decreases from $118$ channels to the $m^{\prime}$ projected channels.

\subsubsection{Continuous wavelet decomposition}

To extract the time-scale information of the EEG signals, we utilize continuous wavelet transform (CWT) and project each trial in different channels on $q$ different scale subspaces.
The time series corresponding to subspace $d$ is computed as follows
\begin{equation}
  f_d(t)=\int_{\mathbb{R}}{WT_{\psi}}(d,s)\psi_{d,s}(t)ds,
\end{equation}
where ${WT_{\psi}}(d,s)$ and $\psi_{d,s}(t)$ are the CWT of the signal and dilated and translated of the mother wavelet.

\section{Methods}\label{section_method}

\subsection{LSTM encoder-decoder}

\begin{figure*}[htbp!]
   \centering
   \includegraphics[width = 0.85\textwidth]{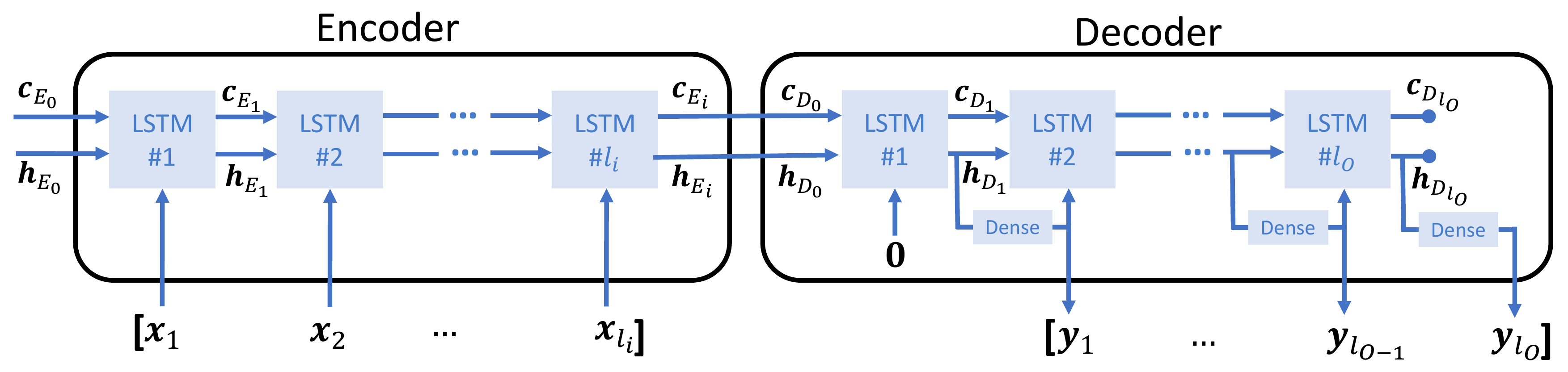}
   \captionsetup{justification=centering}
   \caption{The block diagram of the LSTM encoder-decoder. The encoder generates a fixed-length vector from the entire input sequence of length $\ell_i$ and feeds it to the decoder. The decoder produces the predicted sequence. The ``Dense" block is responsible for resizing the output to fit into the input.}
   \label{ed_fig}
\end{figure*}

Predicting the future time samples enables the BCI system to react quickly to any changes in a user's imagination.
Consequently, the latency of the system is decreased.
We utilize the seq2seq paradigm \cite{sutskever2014sequence} to design our predicting architecture. 
In order to predict the future time samples $\mathbf{Y}$, we must estimate their joint distribution given the past samples $\mathbf{X}$, $P(\mathbf{Y}|\mathbf{X})$.
Assuming future samples are independent, we factorize it according to
\begin{equation}
    P(\mathbf{Y}|\mathbf{X})= \prod^{j}P(\mathbf{y}_j | \mathbf{x}_1,...,\mathbf{x}_{l_i}).
\end{equation}

An LSTM unit \cite{hochreiter1997long} as a variant of recurrent neural network (RNN) can model this conditional probability distribution function efficiently.
To relax the independence assumption, the hidden and cell state at time $t{-}1$, $h_{t-1}$ and $c_{t-1}$, are connected to the hidden and cell state of the next unit at time $t$.
However, this architecture requires that the input and output sequences have the same length.
To get around this limitation, we propose the encoder-decoder (ED) architecture shown in Fig. \ref{ed_fig}. 
The encoder takes input sequences of length $l_i$ and compresses them into context vectors of a fixed-length.
The cell state, the hidden state, and the input of the decoder are initialized using the  final cell state, the final hidden state of the encoder, and a vector $``\mathbf{0}"$, respectively.
In the ED structure shown, the ``Dense" network resizes the output vector to the same length as the input vector.
Finally, $\ell_o$ time samples are generated based on the outputs of the previously computed samples.

\subsection{Modeling the predicting problem as a classification problem}

We solved this problem using regression and classification methods separately, and obtained the best performance with the later method.
Therefore,  we utilize the classification method and quantize our $N$ training trials into $v$ levels. 
For each quantized value, one-hot encoding is applied.
We then use the obtained $(N,\ell_i,m^\prime,q,v)$ and $(N,\ell_o,m^\prime,q,v)$ tensors are as the input and the ground truth data to train $p {=} m^\prime{\times} q$ numbers of EDs. We also use the categorical cross entropy as the loss function according to
\begin{equation}
    \mathcal{Q} = \frac{1}{N\ell_o}\sum_{k=1}^{N}\sum_{i=1}^{\ell_o}\sum_{j=1}^{v} -V_{i,j} \log (\hat{V}_{i,j}),
\label{classcost}
\end{equation}
where $\hat{V}_{i,j}$ and $V_{i,j}$ are the $j$-th element of one-hot encoded vector of the $i$-th sample in the predicted and received sequences, respectively.

The one-hot encoding produces sparse vectors.
We add $\ell_1$--regularization term $\mathcal{Q}_{\ell_1}$ to our loss function $\mathcal{Q}$ to avoid overfitting as follows

\begin{equation}
    \mathcal{Q}_{\ell_1} = \lambda \sum_{i}|w_i|,
\end{equation}
where $\lambda$ and $w_i$'s are the regularization parameter and all trainable weights, respectively. $\ell_1$--regularization functions as a feature selection method and eliminates most of $w_i$ weights.

In the testing phase, when the next time samples corresponding to the predicted ones are received, the similarity $0{\leqslant}\mathrm{S}{\leqslant}1$ between the predicted and received signals is calculated to detect the onset of an MI command, as follows
\begin{equation}
\mathrm{S} = \frac{1}{p\ell_o}\sum_{j=1}^{p} \sum_{i=1}^{\ell_o} (1 - \frac{||\hat{\mathbf{y}}_{i}^{(j)}-\mathbf{y}_{i}^{(j)}||}{||\hat{\mathbf{y}}_{i}^{(j)}||+||\mathbf{y}_{i}^{(j)}||} )\qquad j=1,...,p,
\end{equation}
where $\hat{y}_{i}^{(j)}$ is $i^{th}$ time sample predicted by $j^{th}$ ED and $y_{i}^{(j)}$ is its corresponding received sample.
The predicted sequence is relevant to an MI-task signal if the similarity exceeds a threshold $\mathrm{S}_{th}$. Otherwise, it is regarded as a rest-state signal. 
The threshold is adjusted using 5-fold cross validation in the training phase.
We use the F1-score as the key performance metric to tune our hyperparameters.

\subsection{Error correction algorithm}\label{eca}
As a postprocessing step, we develop a correction algorithm where the label of each segment is modified by looking at its $N_s$ neighbors.
For instance, if we set $N_s\!=\!2$, the label of the current segment is determined by a majority vote on the labels of the preceding, current and the following segment.

\subsection{Performance metrics}\label{metrics}

We use the metrics derived from the confusion matrix \cite{DBLP:journals/corr/abs-2010-16061} to evaluate the performance of our proposed method.
The MI-task and rest-state signals are therefore indicated with a positive and negative label, respectively.
In the following equations, TP, TN, FP, and FN are true positive, true negative, false positive, and false negative, respectively.

The precision metric compares the number of true MI-task signals to all those that have been detected as such:
\begin{equation}
    Precision = \frac{T\!P}{T\!P + F\!P}
\end{equation}

The true positive rate (TPR), also known as recall, is the ratio of MI-task signals correctly detected to all MI-task signals, and the true negative rate (TNR) represents the system ability to stay in the rest state correctly when the user is genuinely in the rest state, as follows
\begin{equation}
    T\!P\!R = \frac{T\!P}{T\!P + F\!N}, T\!N\!R = \frac{T\!N}{T\!N + F\!P}.
\end{equation}

F1-score is defined as the harmonic mean of the precision and the recall so that a high F1-score shows a high level of precision and recall simultaneously:
\begin{equation}
    F1\text{-}score = 2\;\frac{precision \times T\!P\!R}{precision + T\!P\!R}
\end{equation}

\section{Simulation results} \label{section_results}
\subsection{Time information vs. time-scale information}
We use PCA to preserve $70\%$ of the information for dimensionality reduction.
We also train two ED networks with time and time-scale inputs separately to see how time-scale information affects the system performance.
This procedure yields five time series in various projected channels for the time input, and thirty time series in various projected channels and scales for the time-scale inputs with $q{=}6$.
The input and output lengths, the hidden state, the regularization parameter, and the number of epochs are set to $0.5s$, $90$, $0.001$ and $50$, respectively.
The results are presented in Table \ref{timecomparison}.
The results show that the ED trained on time inputs produces a high false-positive rate (FPR).
A possible explanation is that the distribution of the computed similarity for the rest states substantially overlap the distribution of the computed similarity for the MI-task signals.
Hence, many rest states are identified as MI-task signals regardless of the threshold value.
The results also suggest that projecting the time series on different scales improves the system performance significantly for differentiating the MI-task from the rest-state signals.

\vspace{-0.5em}
\begin{table}[hbt!]
\caption{\small Comparison of time and time-scale inputs}\label{timecomparison}
\centering
\begin{tabular}{|c|c|c|c|c|c|c|}
\hline
\rowcolor{DarkGray}\rule{0pt}{1.em}Input & Prec. & TPR & TNR & FPR & FNR & F1 \\[0.2em]
\hline
\rule{0pt}{1.2em} Time only &0.63 &0.935 &0.14 &0.86&0.06 &0.753\\[0.2em]
\hline
\rule{0pt}{1.2em}Time-scale &0.953 &0.892 &0.93 &0.07&0.11 &0.921\\[0.2em]
\hline
\end{tabular}
\end{table}
\begin{table*}[hbt!]
\caption{\small Applying error correction algorithm with $N_s=2$ to the predicted segments}\label{segcor2}
\centering
\resizebox{\textwidth}{!}{%
\begin{tabular}{|c|c|c|c|c|c|c|c|c|c|c|c|c|c|c|c|c|c|c|}
\hline
\rowcolor{DarkGray} \rule{0pt}{1em} \multirow{2}{*}{\shortstack{Hidden \\ states}} & \multicolumn{6}{c|}{$N_s=0$}& \multicolumn{6}{c|}{$N_s=2$}   \\ \cline{2-13} 
\rowcolor{DarkGray} \rule{0pt}{1em}    states  & Prec. & TPR & TNR & FPR & FNR & F1 & Prec. & TPR & TNR & FPR & FNR & F1                          \\[0.1em]\hline
\rule{0pt}{1em}          10          &0.81 &0.86 &0.72 &0.28 &0.14 &0.834&0.88 &0.91 &0.82 &0.18 &0.09 &\textbf{0.895}                            \\[0.1em] \cline{1-13} 
\rule{0pt}{1em}          30          &0.83 &0.92 &0.77 &0.23 &0.08 &0.873 &0.92 &0.97 &0.89 &0.11 &0.03 &\textbf{0.944}                           \\[0.1em] \cline{1-13}
\rule{0pt}{1em}          50          &0.87 &0.96 &0.82 &0.18 &0.04 &0.913& 0.93 &0.98 &0.9 &0.1 &0.02 &\textbf{0.954}                             \\[0.1em] \cline{1-13} 
\rule{0pt}{1em}          70          &0.84 &0.97 &0.79 &0.21 &0.03 &0.9& 0.92 &0.99 &0.89 &0.11 &0.01 &\textbf{0.954}                             \\[0.1em] \cline{1-13} 
\rule{0pt}{1em}          90          &0.91 &0.92 &0.86 &0.14 &0.08 &0.915& 0.96 &0.94 &0.94 &0.06 &0.06 &\textbf{0.95}                            \\[0.1em] \cline{1-13} 
\rule{0pt}{1em}          110         &0.89 &0.96 &0.84 &0.16 &0.04 &0.924&0.95 &0.98 &0.92 &0.08 &0.02 &\textbf{0.965}                            \\[0.1em] \cline{1-13} 
\rule{0pt}{1em}          130         &0.87 &0.97 &0.82 &0.18 &0.03 &0.917&0.94 &0.99 &0.91 &0.09 &0.01 &\textbf{0.964}                            \\[0.1em] \cline{1-13} 
\rule{0pt}{1em}          150         &0.87 &0.95 &0.82 &0.18 &0.05 &0.908& 0.94 &0.97 &0.9 &0.1 &0.03 &\textbf{0.955}                             \\[0.1em] \hline
\rowcolor{Gray}\rule{0pt}{1em}Avg.&0.861 & 0.939& 0.805& 0.195& 0.061 & 0.898&0.93  &0.966 &0.896 &0.104 & 0.034 & \textbf{0.948}                 \\[0.1em] \hline
\end{tabular}%
}
\end{table*}

\vspace{-1em}
\subsection{Evaluating the error correction algorithm}

To see how the correction algorithm affects the system performance, we set the lengths of the input and output, $\ell_i$ and $\ell_o$, to $0.5s$.
To avoid significant delay, we only apply the correction algorithm with $N_s{=}2$.
In addition, we use different numbers of hidden states in this experiment to show the robustness of our proposed model with respect to changes in model parameters.
The results are shown in Table. \ref{segcor2}.
The average F1-score without the correction algorithm is $0.898$.
The correction algorithm improves the average F1-score by $5.6\%$ with $N_s{=}2$.
The correction algorithm also dramatically reduces the FPR and FNR:
In average, the FPR and FNR are reduced by about $46.7\%$ and $44.2\%$, respectively.
Consequently, the correction algorithm can effectively decrease the false alerts.

\subsection{Comparison of the proposed method with the winner of the competition}

A comparison experiment was conducted for the suggested approach and the winner of the BCI competition III \cite{zhang2007algorithm} to evaluate the system performance for detecting the onsets of the MI-task and the rest-state signals.
Here, we employ the first phase of the winner's method that discriminates the rest states from the MI-task signals.
Table \ref{winner1} indicates the results obtained from the winner's method for the inputs with length $0.25s$, $0.5s$, $0.75s$, and $1s$.
The F1-scores obtained with our proposed method are represented in the last column of the table.
The results demonstrate that our proposed technique outperformed the competition winner by improving the average  F1-score by about $26.7\%$.
\begin{table}[hbt!]
\caption{\small Comparison of the winner paper and our proposed method}\label{winner1}
\centering
\begin{tabular}{|c|c|c|c|c|c|c|c|c|}
\hline
\rowcolor{DarkGray}\rule{0pt}{1.em}$\ell_o$ & Prec. & TPR & TNR & FPR & FNR & F1 & ED-F1\\[0.2em]
\hline
\rule{0pt}{1.em}0.25 &0.769 &0.582 &0.73 &0.27&0.42 &0.663& \textbf{0.842}\\[0.2em]
\hline
\rule{0pt}{1.em}0.5 &0.809 &0.585&0.78 &0.22& 0.41&0.679&\textbf{0.887}\\[0.2em]
\hline
\rule{0pt}{1.em}0.75 & 0.862&0.587&0.86 &0.14&0.41 & 0.698& \textbf{0.837}\\[0.2em]
\hline
\rule{0pt}{1.em}1 & 0.869&0.521& 0.88&0.12& 0.48&0.651& \textbf{0.845}\\[0.2em]
\hline
\rowcolor{Gray}\rule{0pt}{1.em}Avg. &0.8270 & 0.569&0.813&0.188 &0.43& 0.673&\textbf{0.853}\\[0.2em]
\hline
\end{tabular}
\end{table}
\vspace{-1em}
\section{Conclusion}\label{conclusion}

We have proposed a self-paced BCI method based on the time series prediction paradigm.
The proposed approach reduces the latency of the BCI system significantly, further allowing the system to be used in real-world applications.
In addition, our suggested method is trained exclusively on MI-task signals and does not need rest-state trials in the training phase for detecting the onsets of MI-tasks interpreted as commands in a continuous EEG signal.
The proposed method is tested on the  available dataset IVc from BCI competition III.
The simulation findings shows that our approach increases the precision, recall, and the TNR, and decreases the FPR and FNR significantly compared to the algorithm proposed by the winner of the competition.
The resulting F1-score is about $26.7\%$ higher than the score achieved by the winner of the competition.

\addtolength{\textheight}{-12cm}  
\bibliographystyle{IEEEtran}
\bibliography{./main.bib}

\end{document}